\documentclass{PoS}
 \usepackage{siunitx}
\usepackage{subcaption}
\usepackage{graphicx}
\usepackage{sidecap}
\title{Laser test with Mini-EUSO}

\ShortTitle{Laser test with Mini-EUSO}

\author{Johannes Eser, \speaker{Viktoria Kungel}, Lawrence Wiencke \\ 
        Colorado School of Mines, Department of Physics, Golden, USA\\
        E-mail: \email{kungel@mines.edu}}

\author{Mario E. Bertaina, Francesca Bisconti\\
        University of Turin, Italy \& INFN Turin, Italy}

\author{Marco Casolino\\
        RIKEN, Wako, Japan \& University of Roma Tor Vergata, Italy \& INFN Roma Tor Vergata, Italy }

\author{for the JEM-EUSO Collaboration\footnote{For collaboration list see PoS(ICRC2019)1177}}
        
\abstract{Mini-EUSO (Extreme Universe Space Observatory) is a small-scale prototype cosmic-ray detector that will measure Earth`s UV emission and other atmospheric phenomena from space.\
It will be placed in the International Space Station (ISS) behind a UV-transparent window looking to the nadir. The launch is planned this year (2019).\
Consisting of a multi-anode photomultiplier (MAPMT) camera and a $25$ cm diameter Fresnel lens system, Mini-EUSO has a \ang{44} field of view (FoV), a $6.5$ km$^2$ spatial resolution on the ground and a $2.5\ \mu$s temporal resolution. \
In principle, Mini-EUSO will be sensitive to extensive air shower (EAS) from cosmic-rays with energies above $10^{21}$ eV. \\
A mobile, steerable UV laser system will be used to test the expected energy threshold and performance of Mini-EUSO.\
The laser system will be driven to remote locations in the Western US and aimed across the field of view of Mini-EUSO when the ISS passes overhead during dark nights. \
It will emit pulsed $355$ nm UV laser light to produce a short speed-of-light track in the detector. \
The brightness of this track will be similar to the track from an EAS resulting from a cosmic-ray of up to $10^{21}$ eV. \
The laser energy is selectable with a maximum of around $90$ mJ per pulse.
The energy calibration factor is stable within $5\ \% $.  
The characteristics of the laser system and Mini-EUSO have been implemented inside the JEM-EUSO OffLine software framework, and laser simulation studies are ongoing to determine the best way to perform a field measurement. 
}

\FullConference{36th International Cosmic Ray Conference -ICRC2019-\\
		July 24th - August 1st, 2019\\
		Madison, WI, U.S.A.}

\begin{document}
\section{Introduction}
\begin{figure}
     \centering
     \begin{subfigure}{.5\textwidth}
    \includegraphics[width=1\linewidth]{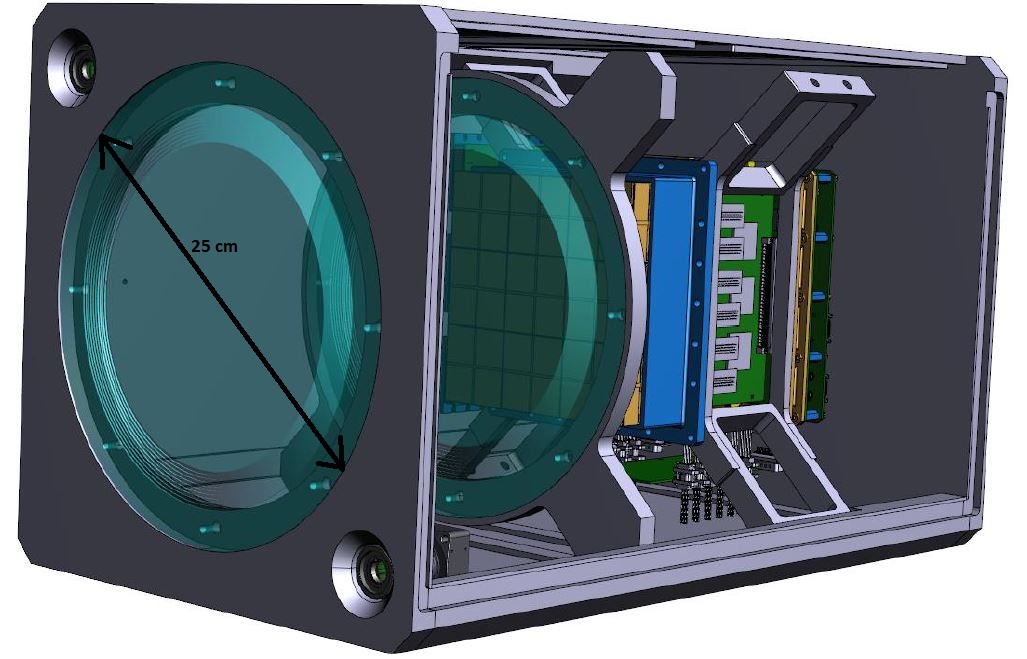}
    \caption{Mini-EUSO design \cite{minipic}.}
    \label{fig:mini}
\end{subfigure} \begin{subfigure}{.48\textwidth}
    \includegraphics[width=1\linewidth]{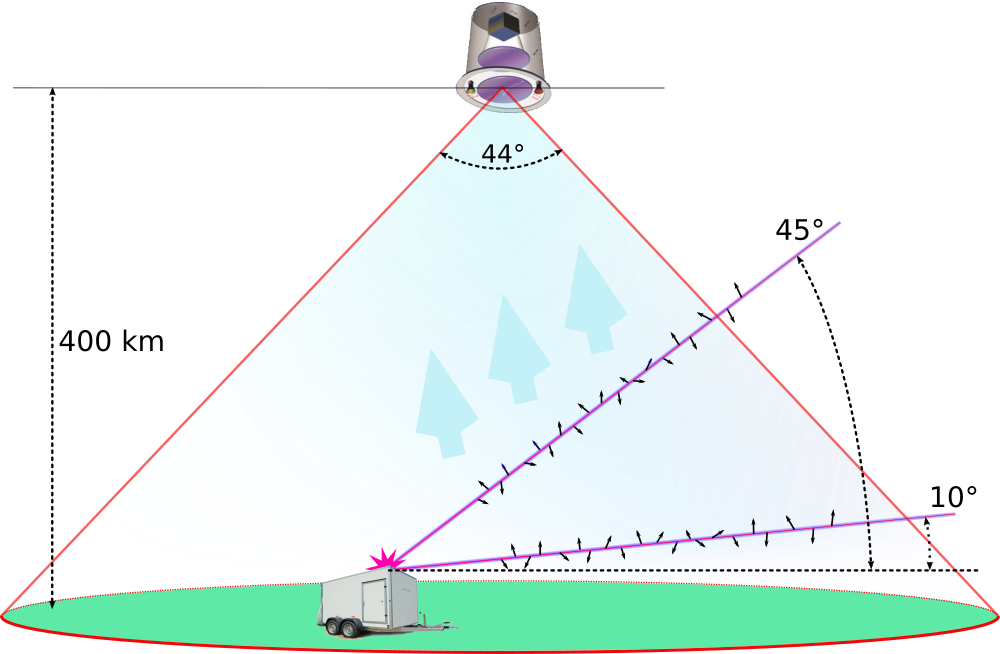}
    \caption{Concept of the laser test. Mini-EUSO is located at 400 km altitude MSL and the laser is aiming in two different angle settings. The Scattered UV light is in the FoV of Mini-EUSO}
    \label{fig:idea}
    \end{subfigure}
	\caption{Mini-EUSO (\ref{fig:mini}) and the configuration of a laser fired in its field of view (FoV).}
    \label{fig:iidea}
\end{figure}
Mini-EUSO is one of the missions of the Joint Experimental Missions for the Extreme Universe Space Observatory (JEM-EUSO) \cite{JEM, EUSOII, Fenu:2017xct, EUSO, Adams:2017fjh}. 
It is a high-speed, multi-wavelength telescope with a 25 cm diameter aperture and a prototype of a space-based ultra-high-energy cosmic-ray detector.
Placed inside the International Space Station (ISS) at 400 km MSL behind a UV-transparent window in nadir direction \cite{MINI}.
Mini-EUSO has a multi-trigger design and thus, three different time resolutions, providing the opportunity to measure various UV light signals like Earth's UV emission, and emissions from atmospheric sources, meteroides or bioluminescence \cite{UV, airglow, meteor}.
Furthermore, it is also capable of detecting space debris.
Mini-EUSO is sensitive to extensive air showers with an energy over $10^{21}$ eV \cite{camera}.
Since the expected rate is too low to catch cosmic-rays with this instrument, a mobile high-power UV laser system on the ground will be used to simulate the optical signature of extensive air showers between $10^{18.5}$ to $10^{21}$ eV \cite{wiencke}.
If successful, this will be the first observation of an artificial speed-of-light track with a fluorescence detector from space.\\

\section{Mini-EUSO}
Mini-EUSO is raising the Technology Readiness Level (TRL) of the JEM-EUSO instrumentation.
One of the technological objectives is the use of Fresnel lenses in space \cite{fresnel, fresnel2, fresnel3}.
The lenses are 10 mm thick, and they have a 25 cm diameter, which is quite thin and suits well to the needs and requirements of a space based experiment by being light and compact. 
Spaced-based activities are regulated within certain norms and comply with international and national standards.
The field of view (FoV) is \ang{44}$\times$\ang{44}.
The $30$-kg fluorescence detector, Mini-EUSO, is depicted in Fig. \ref{fig:mini} and needs $30$ W to operate \cite{30W}.
The camera consists of 36 multi-anode photomultiplier tubes (MAPMT \cite{30kg}).
A MAPMT is approximately $23$ mm$^2$ with a 8$\times$8 pixel array able to count single photoelectrons. 
The full photo detector module (PDM) has 2304 pixel.
Each pixel of Mini-EUSO corresponds to 6.5 km$^2$ on the ground.
The temporal highest-resolution is $2.5\ \mu$s, which is the gate time unit (GTU) \cite{kilo}.
\section{Laser system}
Housed in a customized trailer, a calibrated pulsed 355 nm frequency-tripled YAG laser, has an energy range from $200\ \mu$J to 90 mJ \cite{GLS}.
Aiming across the field of view of Mini-EUSO, the laser light propagates through the atmosphere.
Some of the light scattered out of the beam and reaches the detector's aperture. 
In Fig. \ref{fig:idea}, the UV-laser light is fired in two different setting angles: \ang{10} and \ang{45} above horizon.
The laser is mounted on an optical table allowing a leveling of better than $\pm$\ang{0.024}.
That table is also mechanically isolated from the trailer to minimize vibrations during operation.
The laser head and the optics components are inside an optical enclosure. 
Inside, there is an energy probe connected to a radiometer, that measures a fraction of the laser energy and monitor changes shot-by-shot.
The laser beam leaves the trailer through a UV window at the end of a beam steering periscope.
The laser beam can reach any direction above the horizon with a pointing accuracy higher than $\pm$\ang{0.1}.
Due to the measurements of two energy probes, we can calculate a calibration factor for a relative shot-by-shot measurment.
\section{Laser calibration}
There is a standard procedure developed to calibrate the energy of the laser.
The energy calibration runs automatically at each setting in $5\ \% $ attenuation steps over the full energy attenuation range of the laser and fires (and measures) 50 shots.
The calibration factor is stable for the full energy range of the laser within $5\ \%$.
\begin{SCfigure}
     \centering
    \includegraphics[width=.6\linewidth]{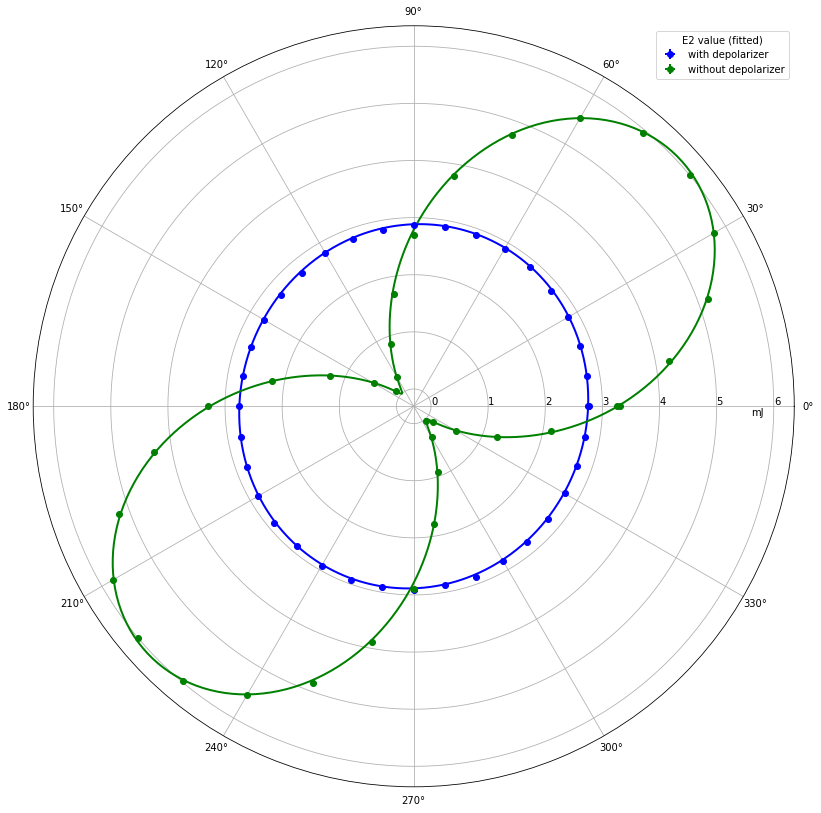}
	\caption{Measured energy of the laser beam at the UV window via a polarization cube with depolarized (blue) and linearly polarized light (green). The laser intensity was damped. The energy distribution for each set of 50 shots is consistent, due to a stable energy calibration factor.}
    \label{fig:pol} 
\end{SCfigure}
As a second calibration step, a polarization calibration via a polarization cube will determine whether the laser beam is polarized or not.
Fig. \ref{fig:pol} shows a polar plot with the energy in mJ on the radial axis for different cube angles.
If the light is completly randomly polarized, we expect a circle, because the randomized beam scatters isotropically in azimuth about the beam axis.
Our measurements (blue dots) are on the circle (fitted blue line), meaning that the depolarizing optics used is adjusted correctly.
If the light is linearly polarized, the measurement (green dots) shows the energy of the beam leaving the UV window on the radial axis in mJ (fitted green line).
The polarized light allows maximum energy that is more than twice as high as the mean energy of the unpolarized.
When operating in linear mode, there are regions that have no light.
Also, we have to take into account, that the steering additionally rotates the linear polarization.
Accurate alignment of the beam is crucial to not miss the signal due to those gaps.\\

\section{Laser simulation}
\begin{figure}
     \centering
     \begin{subfigure}{1.\textwidth}
	     \includegraphics[width=0.5\linewidth]{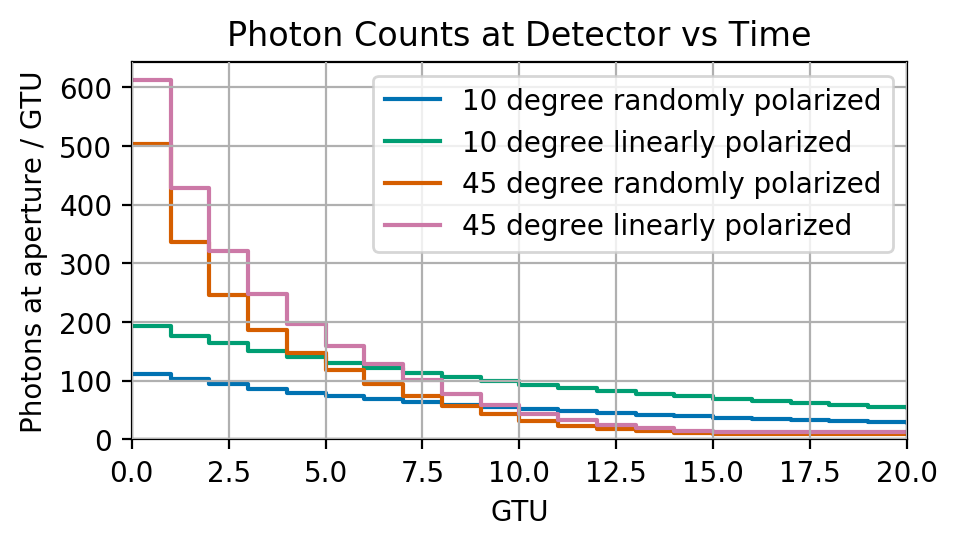}
	     \includegraphics[width=0.5\linewidth]{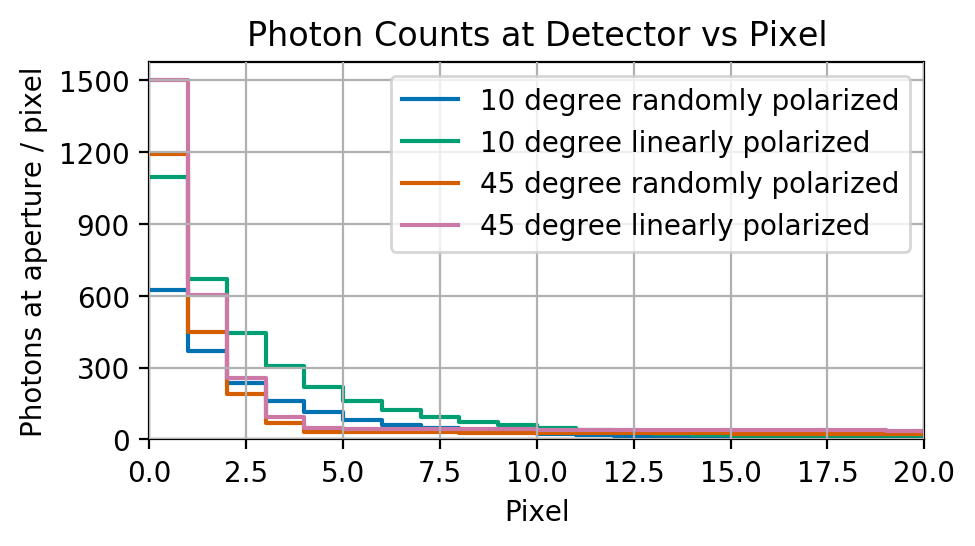}
\end{subfigure}
	\caption{A 355 nm laser simulation with photons hitting the aperture of Mini-EUSO \cite{Austin}. The beam energy used is 90 mJ. Time bins are $2.5\ \mu$s (1 GTU).}
    \label{fig:sim}
\end{figure}

\begin{figure}
     \centering
  \begin{subfigure}{1\textwidth}
\includegraphics[width=0.5\linewidth]{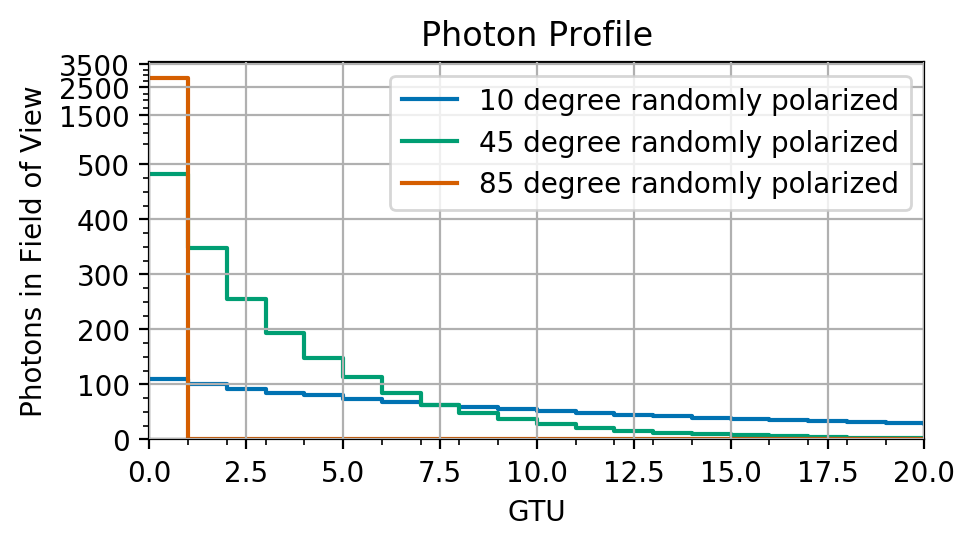}
\includegraphics[width=0.5\linewidth]{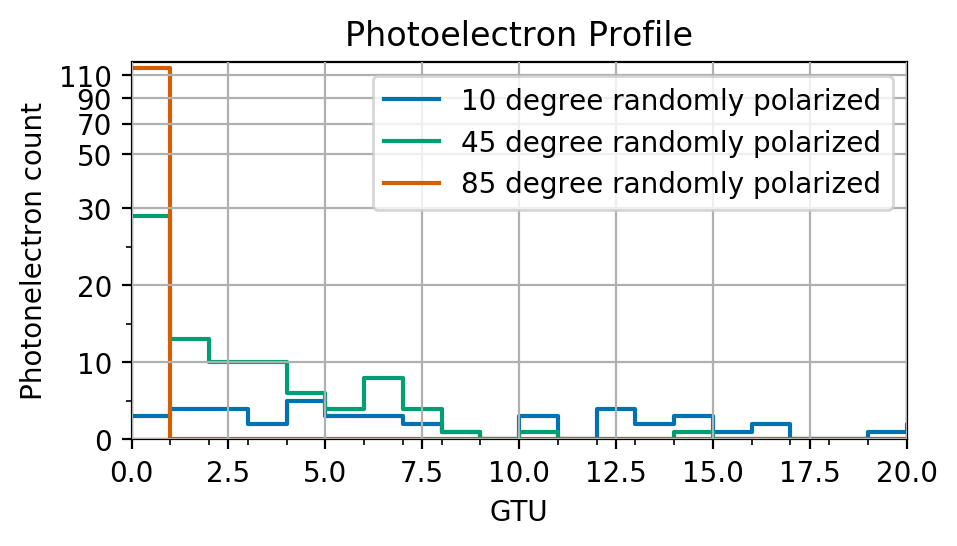}
\end{subfigure} 
\caption{A 355 nm laser simulation including the signal response of Mini-EUSO. The beam energy used is 90 mJ. No background was added to analyze the signal dependencies of different setting angles. The y-axis steps are customized for a better representation.}
\label{fig:lsim}
\end{figure}

The expected illumination of the aperture can be simulated, including taking the atmospheric attenuation into account.
This simulation is shown in Fig. \ref{fig:sim}.
For a laser pointing direction \ang{10} and \ang{45} above the horizon, the expected photon counts per pixel and per time bin are depicted (time bin = 1 GTU).
A laser energy of 90 mJ was used.
Compared to \ang{10}, the photon counts for \ang{45} increase while the number of illuminated pixels decrease.
A simulation for linearly polarized light for \ang{10} and \ang{45} over the horizon, that are constructively aligned to the detector's aperture is also included in Fig. \ref{fig:sim}.
More photon counts in a pixel or per time bin (in GTU) arrive at aperture with polarized light in constructive alignment of the laser beam.
After 10 GTU the signal is almost gone.
Overall we can say, that for higher inclination angles, the track is shorter and so the light is seen by less pixel in less time, but the photon count grows in one pixel.\\
The signal response of Mini-EUSO was simulated with the EUSO-OffLine framework adopted from Auger-OffLine \cite{Offline, JOffline}.
The implementation was tested and is in agreement with laboratory measurements of the instrument.
The simulation with a laser beam light source were done with three different elevation angles of the laser beam, i.e. \ang{85}, \ang{45} and \ang{10}.
The laser is positioned below Mini-EUSO.
It is set to maximal energy of 90 mJ and a 355 nm wavelength.
In Fig. \ref{fig:lsim}, traces of photoelectrons per pixel per GTU are shown.
Closer to the optical axis of the detector, almost vertical at elevation angle \ang{85}, the beam reaches its highest possible intensity, but, the signal length is shorter with less than 1 GTU.
At \ang{10} close to the horizon, the signal is asymmetrically spread and flattened, because of the small signal to noise ratio.
For \ang{45} the signal fades away after 10 GTU as expected and discussed before.
In summary, with higher inclination angle the intensity seen by the detector is higher but the signal is shorter, only approximately 1 GTU at \ang{85}. 
For horizontal tracks, the detector can only see faint scattered light.
\\
\\
\newpage
\section{Planned field campaign}
\begin{figure}
    \centering
    \includegraphics[width=1.\linewidth]{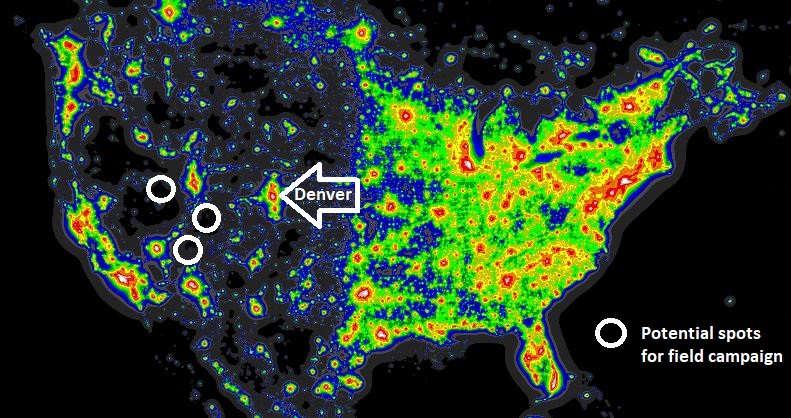}
    \caption{Night light map of the USA. The circles show possible locations for the field campaign \cite{DARK}.}
    \label{fig:dark}
\end{figure}

Because of a risk of light pollution, that can thwart the observation of the UV light, an area with very little background is needed.
The Mini-EUSOs data selection is based on a multi-level trigger logic. 
The data selection works over different timescales \cite{Belov:2017ksp, trigga, triggaalg, triggaalgII}.
The level 1 trigger considers a signal 8 $\sigma$ above background in a timescale of 20 $\mu $s (this is the time for light to cross one pixel) with a $2.5\ \mu$s time resolution.
The level 2 trigger has a much higher integration time scale and thus, receives the integration of 320 $\mu $s (128 GTU = 1 GTU$_{L2}$) from the level 1 trigger.
The different time scales and resolutions refer to different events to be captured and in order to set the background level.
The input of the summation over pixel counts of 1 GTU, will trigger different events/data in different time resolutions after corresponding threshold calculations.
There is a continuous sampling of data (level 3) integrated over 40.96 ms (128 GTU$_{L2}$ = 1 GTU$_{L3}$). 
It stores and sets the background for the level 2 events.
After 128 GTU$_{L3}$ or 5.24 s, all data from the three levels are stored \cite{data}.
The trigger threshold of the level 1 and level 2 triggers are set to an expected signal rate of less than 1 event per 5.24 s. \\
Hence, a very dark site is needed to perform field measurements.
In Fig. \ref{fig:dark} we consider some appropriate spots west of Denver, close to Golden (Colorado), where the light system is located.
Best suited spots are in Utah or Nevada.
Going west is preferred to avoid the cool-down period which would occur if Mini-EUSO was to pass over the Denver metropolitan area.
 \\
The laser will be fired on a moonless night with a clear night sky while the ISS passes overhead.
A field campaign takes almost a week with final preparation on site, set up, operation and return.
It is planned after the launch in August, approximately October 2019. 
The laser sequence can only be fired once, due to the shift of the ISS overflight orbit with each pass over the equator.
Therefore, for each run, an estimate of the ISS orbit and date of overflight as well as a coordination with the astronaut must be made, to guarantuee that the instrument is operational during the overflight. \\

\footnotesize{{\bf Acknowledgment:}{

This work was partially supported

by NASA grants NNX13AH55G, 80NSSC18K0477,

the French Space Agency (CNES),

the Italian Space Agency (ASI),

the Basic Science Interdisciplinary Research Projects of RIKEN and JSPS KAKENHI. 

We also acknowledge the invaluable contributions of the administrative and technical staffs at our home institutions.

}}

\end{document}